%% file: main.tex
\documentclass[preprint,authoryear,12pt]{elsarticle}

\usepackage{booktabs}
\usepackage{graphicx}
\usepackage{xurl}
\usepackage{siunitx}
\usepackage{amsmath}
\usepackage{tikz}
\usetikzlibrary{positioning,arrows.meta,calc,fit,backgrounds}
\biboptions{authoryear}

\journal{Telecommunications Policy}

\makeatletter
\def\ps@pprintTitle{%
  \let\@oddhead\@empty
  \let\@evenhead\@empty
  \let\@oddfoot\@empty
  \let\@evenfoot\@oddfoot}
\makeatother

\begin{document}

\begin{frontmatter}

\title{The Towers Were Standing: A Cause Decomposition of Cellular
       Outages During Hurricane Helene}

%% =========================================================================
%% CHECK BEFORE SUBMISSION: every address below is a placeholder.
%%
%% Editorial Manager e-mails each listed co-author to confirm authorship, so
%% all four must work at submission time, not just the corresponding one.
%%
%% On institutional vs. personal addresses: elsarticle accepts both in one
%% \ead, comma-separated, and that is the arrangement used here for the authors
%% who expect to lose their university mailbox --
%%
%%     \ead{name@gmu.edu, name@gmail.com}
%%
%% which prints as "name@gmu.edu, name@gmail.com (Author Name)". Listing the
%% institutional address first keeps the affiliation verifiable; the second
%% address is what still works after the first is deactivated. An ORCID iD,
%% entered separately in Editorial Manager, is the durable identifier and does
%% not depend on either mailbox.
%% =========================================================================
\author[usm]{Oluseyi Olukola\corref{cor1}}
\ead{Oluseyi.olukola@usm.edu}

\author[gmu]{Oare Danielle Addeh}
\ead{oaddeh@gmu.edu}

\author[gmu]{Esther Abiodun Konan}
\ead{ekonan@gmu.edu}

\author[usm]{Nick Rahimi}
\ead{Nick.rahimi@usm.edu}

\cortext[cor1]{Corresponding author.}

\affiliation[usm]{organization={School of Computing Sciences and Computer
                                Engineering, University of Southern
                                Mississippi},
                  city={Hattiesburg},
                  state={MS},
                  postcode={39406},
                  country={United States}}

\affiliation[gmu]{organization={Department of Communication, College of
                                Humanities and Social Sciences, George Mason
                                University},
                  city={Fairfax},
                  state={VA},
                  postcode={22030},
                  country={United States}}

\begin{abstract}
Hurricane Helene produced the largest absolute cell-site outage in the public
FCC record, peaking at \num{4562} sites. The conventional model is physical:
towers destroyed. Helene did destroy over \num{1700} miles of fibre, but
almost none of it was cell sites. We present the first \emph{cause-decomposed}
study of the FCC's Disaster Information Reporting System, reconstructing
\num{80} state-days and \num{580} county-days from 24 daily filings by two
reconciled independent extractions. Damage to cell sites is negligible:
\SI{1.1}{\percent} of attributed cell-site-days across six states, at most
\SI{3.8}{\percent} anywhere. The sites were standing. What took them out
divides by terrain: pooled, power dominates at \SI{63.2}{\percent}, but in
mountainous North Carolina severed \emph{transport} (backhaul) reaches
\SI{52.2}{\percent} against \SI{47.3}{\percent}, and in Tennessee
\SI{69.9}{\percent}. North Carolina's transport share rises from
\SI{7.0}{\percent} to \SI{85.0}{\percent} across the event ($\rho = 0.92$).
Seventeen days after landfall, on 15 October, 47 sites lost transport across
six contiguous North Carolina counties with no rainfall, no power loss, no
damage, and recovery by the next report. Independent active-probe measurement
corroborates it: responsive \texttt{/24}s fall \SI{1.02}{\percent} for twelve
hours while Tennessee stays flat. We release the dataset. Backup power is the
standard resilience investment; here it addresses the smaller half of the
problem.
\end{abstract}

\begin{keyword}
network outages \sep disaster resilience \sep cellular networks \sep
backhaul \sep network measurement \sep FCC DIRS \sep telecommunications
regulation
\end{keyword}

\end{frontmatter}

\section{Introduction}
\label{sec:intro}

On 26 September 2024 Hurricane Helene made landfall in Florida and tracked
inland across Georgia, the Carolinas, Tennessee and Virginia. In the mountains
of western North Carolina it produced catastrophic flooding, and with it the
largest absolute cell-site outage in the public DIRS record: \num{4562} cell
sites out of service on 28 September, across the six states reporting that day
(FL, GA, NC, SC, TN, VA)~\citep{fcc_dirs_helene}.\footnote{We say ``largest in
the DIRS record'' rather than ``largest ever'' because the comparison does not
extend further. Hurricane Irma reached \num{4370} sites out on 11 September
2017, only \SI{4}{\percent} below Helene~\citep{fcc_dirs_irma}, and for
Superstorm Sandy the Commission published only that more than a quarter of cell
sites in 158 counties across ten states and the District of Columbia were
disabled at peak, with no absolute count~\citep{fcc2013resiliency}. DIRS counts
also scale with the county set the FCC designates, which differs by event:
Helene's reporting area held \num{37717} sites (\SI{12.1}{\percent} out)
against Irma's \num{16352} (\SI{26.7}{\percent}). Proportionally the largest
DIRS outage remains Hurricane Maria, at \SI{95.6}{\percent} of Puerto Rico's
\num{1789} sites on 23 September 2017~\citep{fcc_dirs_maria}. For scale within
the mainland record, Hurricane Ida peaked at \num{1824} sites out across
Alabama, Louisiana and Mississippi on 30 August 2021~\citep{fcc_dirs_ida}.}
Over \num{1700}
miles of fibre-optic cable were destroyed, 19 North Carolina counties were left
technologically isolated, and, on 28 September, 17 Public Safety Answering
Points lost the ability to receive 911 calls~\citep{domprep2024helene}.
Clinicians at a rural hospital in Henderson County, one of the six counties
in the October event we isolate in \S\ref{sec:res:event}, describe treating
patients without electronic records, laboratory systems or their usual
communication tools for the duration~\citep{gamboa2026helene}.

The public and engineering narrative of such an event is physical: towers fall,
equipment floods, and restoration is understood as reconstruction. That framing shapes what gets funded, what gets hardened, and
what researchers build: if the problem is destruction, the answer is sturdier
infrastructure and faster rebuilding.

\paragraph{What this paper does and does not claim}
We do not claim Helene destroyed little. It destroyed a great deal, and the
\num{1700} miles of severed fibre are themselves physical damage of the most
literal kind. Our claim is about \emph{where on the network graph} that damage
landed, and it is a claim the reporting instrument makes precise. DIRS asks
providers to attribute each out-of-service cell site to damage \emph{at the
site}, to loss of \emph{transport} (the backhaul linking that site to the
network core), or to loss of commercial \emph{power}. Read in those terms, the
answer is unambiguous: across six states, damage at the site accounts for
\SI{1.1}{\percent} of attributed cell-site-days, and never exceeds
\SI{3.8}{\percent} in any state. The radio access equipment was overwhelmingly
intact; what had failed, in the mountains, was the transport link between that
equipment and the network core.

That distinction is not bookkeeping. A site down for lack of power and a site
down for lack of backhaul present identically to a subscriber, but they are
different engineering problems, with different restoration timelines, different
mitigations, and (as we show) different temporal behaviour. Backup power,
the standard hardening investment, addresses only one of them.

\paragraph{Scope}
The balance between the two shifts with terrain, and the paper scopes its
claims accordingly. Pooled across all six reporting states, commercial power remains
the larger cause (\SI{63.2}{\percent} of attributed site-days against
\SI{35.7}{\percent} transport). It is in the mountainous inland states (North
Carolina and Tennessee, where fibre follows a small number of valley routes)
that transport overtakes it. The finding that generalises across our whole
sample is the negligibility of damage at the site; the transport-dominance
finding is scoped to constrained-route terrain, which is also where the
outage was largest and longest.

\subsection{Background: the measurement instrument}
\label{sec:background}

DIRS is a voluntary reporting system the FCC activates for major
disasters~\citep{fcc_dirs_program}. Communications providers file daily status
for a defined disaster area; the Public Safety and Homeland Security Bureau
publishes a daily aggregate as a PDF, stating the instant its data describe
(09:00 EDT on the report date, for the Helene activation).

Crucially for this work, the published tables break each day's cell-site
outages into three causes (sites out due to \emph{damage}, due to
\emph{transport}, and due to \emph{power}) plus a count of sites operating on
backup power. This decomposition is the measurement instrument we exploit.
Prior work has used FCC hurricane reports as a source of outage \emph{counts}
(\S\ref{sec:related}); to our knowledge the cause columns have not previously
been analysed.

DIRS's mandatory sibling, the Network Outage Reporting System (NORS), collects
richer per-incident detail but is presumed confidential under
47~C.F.R.\ \S4.2 \citep{cfr47part4,fcc_nors}, and is therefore unavailable for
research without a confidentiality process~\citep{claffy2022challenges}.

The Commission has recently modernised the regime. Its Third Report and Order
on Resilient Networks, adopted 20 May 2026, expands the ability of DIRS filers
to submit geospatial information voluntarily, including cell-site locations,
and harmonises transport-facility reporting to optical-carrier
circuits~\citep{fcc2026resilient,fcc2026resilientfr}. Our
results speak directly to that proceeding.

\subsection{Contributions}

\begin{itemize}
  \item \textbf{A dataset and a reconciliation method} (\S\ref{sec:method}).
    A per-cause, per-county daily record of Helene's cellular outage
    (\num{80} state-days and \num{580} county-days), reconciled by the
    cross-table method of \S\ref{sec:method:extraction}.
  \item \textbf{Cause decomposition} (\S\ref{sec:res:decomposition}).
    Damage at the site is negligible in all six states; transport exceeds power
    only in the two mountainous ones. We report both denominators and the
    unattributed residual.
  \item \textbf{The temporal inversion} (\S\ref{sec:res:temporal}). North
    Carolina's transport share rises near-monotonically across the event
    ($\rho = 0.92$), reaching \SI{80.9}{\percent} of attributed outages in the
    recovery tail. We test the trend against both serial correlation and the
    mid-event rescoping of the reporting area.
  \item \textbf{An isolating event, externally corroborated}
    (\S\ref{sec:res:event}). A 15 October transport failure across six
    contiguous counties, with weather, power, damage and rescoping eliminated
    against primary sources, and the timing confirmed by independent
    active-probe measurement.
\end{itemize}

%% ===========================================================================
\section{Related Work}
\label{sec:related}

\paragraph{Outage measurement from the outside}
A substantial literature infers outages from active and passive Internet
measurement. Dainotti et al.\ characterised politically-motivated outages from
BGP, network-telescope traffic and
traceroute~\citep{dainotti2011outages,dainotti2014outages}; Trinocular
adaptively probes a sample of addresses within each responsive \texttt{/24} and
infers block-level reachability outages~\citep{quan2013trinocular}; and Bischof
et al.\ characterised Internet outages and shutdowns at
scale~\citep{bischof2023destination}.

These approaches can attribute cause, but only at a coarse and largely
political granularity: Bischof et al.\ separate government-ordered shutdowns
from spontaneous outages, and Dainotti et al.\ identify censorship. None
reports the operating cause an operator would file (damage, transport or
power) and, for our purposes, the platforms are least able to see exactly
the networks we care about: Bischof et al.\ note that active probing of
publicly routable IPv4 space has limited visibility into NAT-heavy mobile
networks. Our work is complementary in both directions: DIRS reports operating
cause directly for cellular, at the price of covering only declared disasters
and only participating providers, and we use an external platform to
corroborate DIRS in \S\ref{sec:res:ioda}.

\paragraph{Disasters and physical infrastructure}
Cho et al.\ measured ISP impact of the 2011 Tōhoku
earthquake~\citep{cho2011japan}; Heidemann et al.\ analysed Hurricane
Sandy~\citep{heidemann2012sandy}; Padmanabhan et al.\ linked weather events to
residential link failures at scale using
ThunderPing~\citep{padmanabhan2019weather}; Durairajan et al.\ assessed long-run
sea-level-rise risk to US Internet
infrastructure~\citep{durairajan2018lightsout}. Maitland and Peha studied Puerto
Rico's post-Maria restoration from a policy
standpoint~\citep{maitland2018puertorico}. None decomposes a \emph{cellular}
outage record by reported cause. Transport and power are occasionally
separated in passing (Cho et al.\ describe severed backbone circuits and,
separately, a data centre losing mains power) but qualitatively, for a
single operator, and not as a decomposition of the outage.

\paragraph{Cellular resilience engineering}
\citet{booker2010cellular} and \citet{griffith2015resiliency} model base-station survival under hurricanes;
Malandrino and Chiasserini consider disaster impact on wireless
communication~\citep{malandrino2017disasters}. Closest to our finding, Yang et
al.\ analyse backhaul failure and resilient backhaul design using
operator-internal data~\citep{yang2017backhaul}. This literature largely models
\emph{physical} survival, which our results suggest is the smaller problem.
Backhaul is not absent from it (Booker et al.\ simulate the resilience
benefit of fibre-ring and mesh topologies) but it is reached either by
simulation or, in Yang et al.'s case, from proprietary operational data
unavailable to the community, and in neither case as an observed share of a
real outage.

\paragraph{FCC outage data outside networking}
\label{sec:related:fcc}
FCC outage reporting has been used before, but as a source of outage counts
rather than causes, and largely outside the networking literature. Reed and
Wang built cell-tower fragility and recovery curves for Hurricanes Harvey and
Irma from FCC cell-tower outage reports alongside utility and Department of
Energy data~\citep{reed2018numerical}. Du models county-level
telecommunications outages across ten hurricanes and eight states, using outage
magnitude as the prediction target and power outage and demographics as
predictors~\citep{du2024telecomoutage}.\footnote{Du does not name the outage
source in the abstract and the article is paywalled; we therefore describe what
it models rather than assert which FCC series it draws on.} Kuhn analysed the
pre-NORS FCC outage-reporting regime for the public switched telephone network,
and did decompose those outages by cause (into human error, natural events,
hardware, software, overload and vandalism) though that taxonomy has no
transport or commercial-power category and does not concern
cellular~\citep{kuhn1997pstn}. claffy and Clark survey NORS and DIRS
as public-interest measurement instruments and document their access
restrictions~\citep{claffy2022challenges}. Recent Congressional analysis
describes both systems, the restoration of cellular service after Helene, and
their data limitations~\citep{crs2025cellularoutage,gallagher2024helene}.

In all of this work, an outage is a scalar.

\paragraph{The cause columns are an unused resource}
Feeny et al.\ illustrate this. They reconstruct cellular coverage under natural
hazards from public data and, needing failure causes, model structural damage
and power loss from first principles, scoping backhaul out of the
analysis~\citep{feeny2026quantitative}. That is a reasonable design given the
sources they draw on. It also indicates that a 2026 paper reaching for
cause-resolved cellular outage did not have the DIRS decomposition in view,
which is better evidence that the columns are unexploited than an absence of
citations could give.

We therefore make the narrow claim, and only it: this is the first
study to exploit the DIRS cause decomposition, and hence the first to separate
transport from power in a public cellular outage record. We do not claim to be
the first to use DIRS.

%% ===========================================================================
\section{Methodology}
\label{sec:method}

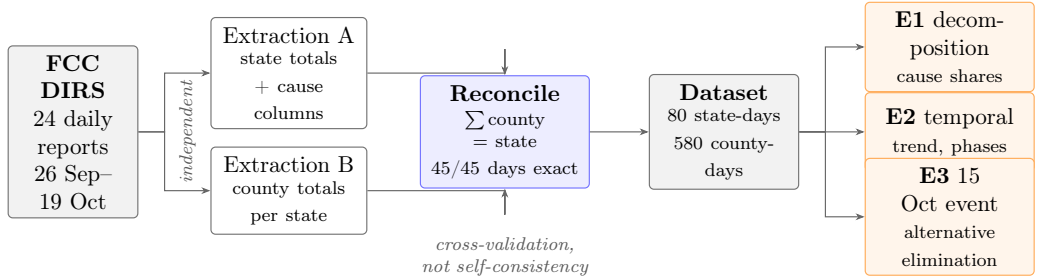
\begin{figure}[htbp]
\centering
\resizebox{\linewidth}{!}{\input{figures/methodology.tex}}
\caption{Measurement pipeline. Two independent extractions (one targeting
state-level tables with their cause columns, one targeting county-level tables)
are reconciled against each other before any analysis. The two read
\emph{disjoint} tables, so the reconciliation is a cross-table identity rather
than a repeat of the same read. Agreement is exact on all \num{45} reconcilable
state-days: \num{23} in North Carolina and \num{22} in Tennessee.}
\label{fig:method}
\end{figure}

Figure~\ref{fig:method} summarises the pipeline. Its central feature is the
fork: two extractions are performed independently from the same source
documents and compared only afterwards, so agreement between them is evidence
rather than tautology.

\subsection{Data acquisition}

We retrieved every published DIRS Communications Status Report for Helene
directly from the Commission's document
server~\citep{fcc_dirs_helene}. Twenty-four reports exist, one for each day from
26 September to 19 October 2024 inclusive, including the weekend of 12--13
October. The 26 September report predates state-level cause reporting and is
used only for county baselines, giving 23 days of analysable cause data. Each report states
the instant it describes (09:00 EDT on the report date), so the
series is a sequence of point samples at a fixed daily offset, not a set of
daily aggregates. \S\ref{sec:res:ioda} depends on that.

Table~\ref{tab:fields} lists the fields extracted from each report.

\begin{table}[htbp]
\caption{Fields extracted from each daily DIRS report. The three cause columns
are the measurement instrument this study depends on.}
\label{tab:fields}
\small
\begin{tabular}{lll}
\toprule
Level & Field & Used for \\
\midrule
State  & sites served, sites out         & denominators \\
State  & out due to \emph{damage}        & E1, E2, E3 \\
State  & out due to \emph{transport}     & E1, E2, E3 \\
State  & out due to \emph{power}         & E1, E2, E3 \\
State  & sites on backup power           & E3 \\
County & sites served, sites out         & reconciliation, E3 \\
Area   & subscribers out (cable/wireline)& E3 \\
Area   & PSAP status                     & context \\
\bottomrule
\end{tabular}
\end{table}

\subsection{Extraction and reconciliation}
\label{sec:method:extraction}

The FCC publishes these tables as PDFs whose text layer collapses adjacent
numeric columns, so naive extraction is unreliable. We therefore performed two
independent extractions (Extraction~A targeting state-level tables and their
cause columns, Extraction~B targeting county-level tables) and cross-validated
them.

The independence that matters here is not two runs of the same parser. A and B
read \emph{different tables}, on different pages, with different column layouts,
and the check between them is an accounting identity that the source documents
never print: for every day, the sum of the county rows must equal the
independently extracted state row. A shared tooling bug would have to produce
compensating errors in two differently-shaped tables to survive that test.
The identity holds on every day it can be checked. Across all \num{45}
state-days for which both a county table and an independently extracted state
row exist (\num{23} in North Carolina and \num{22} in Tennessee), the sum
of the county rows reproduces the state totals \emph{exactly}, in both
sites-served and sites-out. The one county table that cannot be checked this way
is North Carolina's on 26 September, whose report predates state-level cause
reporting; it is excluded rather than counted.

Where a row's text layer was corrupt, the value was recovered by solving against
the report's own printed percentage and column totals, and accepted only when
the solution was unique; one such row (Buncombe, 29 September) occurred and was
resolved and verified two ways. No value is interpolated, and days without a
published report are absent rather than filled.

\subsection{Dataset composition}

Table~\ref{tab:dataset} gives the resulting dataset.

\begin{table}[htbp]
\caption{Composition of the reconstructed dataset. Panels are unbalanced
because states and counties enter and leave the FCC's disaster area as the
event evolves.}
\label{tab:dataset}
\small
\begin{tabular}{llrrr}
\toprule
Level & Scope & Days & Units & Records \\
\midrule
State  & FL, GA, NC, SC, TN, VA & 23 & 6 states   & \num{80} \\
County & North Carolina         & 24 & 21 counties& \num{420} \\
County & Tennessee              & 22 & 8 counties & \num{160} \\
\midrule
\multicolumn{4}{l}{Total county-days} & \num{580} \\
\bottomrule
\end{tabular}
\end{table}

\subsection{Definitions and derived quantities}
\label{sec:definitions}

\paragraph{Cell-site-day} One site out of service for one reporting day. We
integrate over the event rather than reporting a single snapshot, because
snapshots are dominated by whichever day is chosen.

\paragraph{Two denominators, both reported} The three cause counts do not sum
to the printed outage total. Across North Carolina they account for \num{6373}
of \num{6845} out-of-service site-days, leaving \num{472} (\SI{6.9}{\percent})
unattributed; pooled across six states the residual is \SI{7.6}{\percent}. We
therefore define the transport share against attributed causes,
\[
  s_{\mathrm{transport}} \;=\;
  \frac{n_{\mathrm{transport}}}
       {n_{\mathrm{damage}} + n_{\mathrm{transport}} + n_{\mathrm{power}}},
\]
and report the alternative (transport as a fraction of all out-of-service
site-days) alongside it wherever a headline number appears. For North
Carolina the transport share is \SI{52.2}{\percent} of attributed causes and
\SI{48.6}{\percent} of all out-of-service site-days; these are the same
quantity under two denominators, and should not be confused with power's
\SI{47.3}{\percent}, which is a different cause under the first of them. The
choice of denominator changes no ordering, because both causes are scaled by
the same residual: transport exceeds power either way
(\SI{52.2}{\percent} vs \SI{47.3}{\percent} attributed;
\SI{48.6}{\percent} vs \SI{44.0}{\percent} of all out-of-service site-days).

\paragraph{Within-state shares, because the panel moves} The FCC rescopes
the disaster area as the event evolves, and it does so twice in ways that matter
here: Georgia and South Carolina leave after 7 October, and 12 of North
Carolina's 21 counties leave after 12 October, dropping its served count from
\num{1452} sites to \num{901}. Absolute counts are therefore not comparable
across those boundaries: a naive reading of the raw subscriber series would
show an overnight fall from \num{325348} to \num{84085} on 8 October, which is
Georgia and South Carolina leaving, not recovery.

We consequently compute every trend on within-state shares, never on absolute
counts, and we never compare a count across a rescoping boundary. That
convention removes the arithmetic problem but not the compositional one: if the
departing counties differed systematically in cause mix, a within-state share
could still move for the wrong reason. \S\ref{sec:res:rescope} bounds that
possibility directly, and it is the reason the tail-phase results are
trustworthy.

%% ===========================================================================
\section{Experiment}
\label{sec:experiment}

We run four analyses on the reconstructed dataset.

\paragraph{E1: cause decomposition}
Integrate cell-site-days by cause over the full record, per state and pooled.
Tests whether the outage was predominantly physical, and whether the answer
depends on terrain. Reported in \S\ref{sec:res:decomposition}.

\paragraph{E2: temporal behaviour}
Compute the daily transport share and test it for trend, correcting for the
serial correlation a monotone recovery induces, and separately for the
mid-event change in the reporting panel. Then aggregate into event phases.
Reported in \S\ref{sec:res:temporal} and \S\ref{sec:res:rescope}.

\paragraph{E3: the 15 October event}
An anomaly detected in E2: the North Carolina outage count rose mid-recovery.
We test each candidate explanation (rescoping, weather, power, damage, and a
region-wide disruption) against an independent source, then examine the
county-level geographic signature of the residual. Reported in
\S\ref{sec:res:event}.

\paragraph{E4: external corroboration}
DIRS is self-reported by the providers whose networks failed. We ask whether an
independent instrument saw the 15 October event, using IODA's active-probe
signal with Tennessee (in the same activation, measured the same way) as a
control. Reported in \S\ref{sec:res:ioda}.

%% ===========================================================================
\section{Results}
\label{sec:results}

\subsection{E1: cause decomposition}
\label{sec:res:decomposition}

Table~\ref{tab:decomposition} gives the headline result for North Carolina
under both denominators.

\begin{table}[htbp]
\caption{Cause decomposition of North Carolina cell-site-days out of service
over all 23 reporting days, 27 September -- 19 October 2024, under both
denominators
(\S\ref{sec:definitions}). Transport exceeds power under either.}
\label{tab:decomposition}
\small
\begin{tabular}{lrrr}
\toprule
Cause & Cell-site-days & \% attributed & \% of all out \\
\midrule
Transport (backhaul) & \num{3325} & \SI{52.2}{\percent} & \SI{48.6}{\percent} \\
Power                & \num{3012} & \SI{47.3}{\percent} & \SI{44.0}{\percent} \\
Damage at site       & \num{36}   & \SI{0.56}{\percent} & \SI{0.53}{\percent} \\
\midrule
Total attributed     & \num{6373} & & \SI{93.1}{\percent} \\
Unattributed         & \num{472}  & n/a & \SI{6.9}{\percent} \\
\midrule
Sites out            & \num{6845} & & \\
\bottomrule
\end{tabular}
\end{table}

These figures come from Extraction~A of the pipeline in
Figure~\ref{fig:method}; the county totals that validate them come from
Extraction~B. Every figure in this section satisfies the cross-table identity
of \S\ref{sec:method:extraction}.

Two observations follow. \textbf{Damage at the site is negligible}: at
\SI{0.56}{\percent} of attributed site-days, it is two orders of magnitude
below either other cause. \textbf{Transport exceeds power}, which is the
finding with engineering consequences, because backup power is the standard
resilience investment for cell sites and it addresses only the power share.
Battery capacity does not restore a site whose fibre has been cut.

\paragraph{What the unattributed residual can and cannot do to this}
\SI{6.9}{\percent} of North Carolina site-days, and \SI{7.6}{\percent} pooled,
carry no cause. Reporting that number is not the same as saying what it could
do, so we test both conclusions against it adversarially.

The damage figure is robust in one direction only. \SI{0.56}{\percent} is a
\emph{floor}, not a mean. If ambiguous
filings (a site simultaneously flooded, unpowered and cut off) are more
likely to be left unattributed when the true cause is damage, then damage is
understated. Giving the residual entirely to damage, which is the most hostile
assumption available, raises it to \SI{7.4}{\percent} of North Carolina
out-of-service site-days and \SI{8.6}{\percent} pooled. Even that ceiling leaves
damage the smallest of the three causes everywhere, so the negligibility claim
survives, but it should be stated as ``at most \SI{8.6}{\percent}, and
\SI{1.1}{\percent} on the attributions actually filed'' rather than as
``\SI{1.1}{\percent}''.

The transport-exceeds-power ordering in North Carolina is \emph{not} robust to
the same treatment. Transport leads power by 313 site-days; the residual is
472. Allocating more than \SI{66}{\percent} of the residual to power reverses
the ordering.

Three things bear on whether that allocation is plausible. First, across the 22
North Carolina days with more than 20 sites out, the residual's size shows no
relationship to the power share ($r = -0.01$, $p = 0.96$; Spearman
$\rho = -0.03$, $p = 0.91$), a null result consistent with noise rather than
with a cause-dependent filing convention. Second, the ordering holds under any
allocation preserving the observed proportions, since both causes scale
together. Third, Tennessee's residual is \SI{0.8}{\percent} and cannot move
its \SI{69.9}{\percent} transport share, so the transport-dominance finding
does not rest on the state with the large residual. None of this excludes a
power-skewed residual in North Carolina; it establishes that we see no sign of
one.

\paragraph{How far this generalises}
Table~\ref{tab:states} extends the decomposition to all six reporting states and
gives the pooled figure, which we report because the North Carolina result
alone would misrepresent the event.

\begin{table}[htbp]
\caption{Cause decomposition by state, with the six-state pooled total. Shares
are of attributed causes; the last column gives transport as a fraction of all
out-of-service site-days, and the residual column the unattributed remainder.
Damage never exceeds \SI{3.8}{\percent}. Transport \emph{dominance} over the
whole record is confined to the two mountainous inland states, but the shift
toward transport is not (Table~\ref{tab:phases}).}
\label{tab:states}
\small
\setlength{\tabcolsep}{4pt}
\begin{tabular}{lrrrrrr}
\toprule
State & Days & Out & Transp. & Power & Dmg & Transp. \\
      &      &     &         &       &     & (of out) \\
\midrule
TN & 22 & \num{1058} & \SI{69.9}{\percent} & \SI{30.1}{\percent} & \SI{0.0}{\percent} & \SI{69.4}{\percent}\\
NC & 23 & \num{6845} & \SI{52.2}{\percent} & \SI{47.3}{\percent} & \SI{0.6}{\percent} & \SI{48.6}{\percent}\\
VA &  6 & \num{616}  & \SI{42.1}{\percent} & \SI{56.9}{\percent} & \SI{1.0}{\percent} & \SI{40.6}{\percent}\\
GA & 11 & \num{5707} & \SI{30.9}{\percent} & \SI{68.0}{\percent} & \SI{1.2}{\percent} & \SI{28.4}{\percent}\\
SC & 11 & \num{4905} & \SI{14.4}{\percent} & \SI{84.3}{\percent} & \SI{1.3}{\percent} & \SI{12.9}{\percent}\\
FL &  7 & \num{1282} & \SI{14.1}{\percent} & \SI{82.2}{\percent} & \SI{3.8}{\percent} & \SI{13.3}{\percent}\\
\midrule
\textbf{Pooled} & 23 & \num{20413} & \SI{35.7}{\percent} & \SI{63.2}{\percent} & \SI{1.1}{\percent} & \SI{33.0}{\percent}\\
\bottomrule
\end{tabular}
\end{table}

The pooled figure is the right headline for Helene as a whole: \emph{power was
the larger cause}, at \SI{63.2}{\percent} of attributed site-days against
\SI{35.7}{\percent} for transport. Georgia (\num{5707} site-days) and South
Carolina (\num{4905}) are comparable in magnitude to North Carolina
(\num{6845}) and both fall on the power-dominated side. Any claim that
transport is the dominant cellular failure mode in hurricanes generally is not
supported by these data, and we do not make it.

Two things do survive pooling. The first is the negligibility of damage at the
site: no state exceeds \SI{3.8}{\percent}, and the pooled figure is
\SI{1.1}{\percent}. Whatever took cell sites out of service during Helene, in
every state, it was overwhelmingly not the destruction of the site. The second
is the ordering. Tennessee and North Carolina, the mountainous inland states
where fibre follows a small number of valley routes, show the highest
transport shares, and Florida and South Carolina the lowest. We report the
ordering as an observation rather than a result: six states is a small sample,
the states differ in exposure duration as well as terrain (Virginia contributes
six days, North Carolina 23), and we have no independent measure of route
diversity to test the mechanism against. The scoped claim we do make is that
\emph{in constrained-route terrain the topological failure mode dominates}, and
that this terrain is where Helene's outage was largest and longest-lived.

\subsection{E2: the temporal inversion}
\label{sec:res:temporal}

The aggregate decomposition understates the point, because the mix is not
stationary.

\begin{figure}[htbp]
\centering
\includegraphics[width=\linewidth]{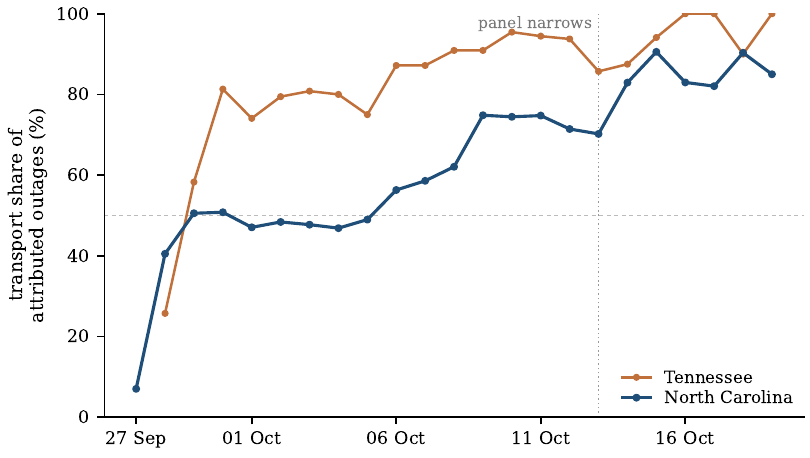}
\caption{Transport share of attributed cell-site outages, by day. Two states
served by different provider mixes show the same inversion.}
\label{fig:inversion}
\end{figure}

Figure~\ref{fig:inversion} plots the daily transport share for North Carolina
and Tennessee, the only two states whose records span the whole activation:
23 and 22 reporting days, against 11 for Georgia and South Carolina and 6 for
Virginia. A daily trend test on six points is not worth running. The phase-level comparison later in this section
(Table~\ref{tab:phases}) covers the shorter records too, and it is where the
generality of the pattern is actually tested. \ref{app:daily} gives
North Carolina's complete daily record (served, out, and all three cause
columns for every reporting day), so that every point in this figure and the
next can be checked against the filings.

The quantity of interest is the size of the shift. North Carolina's transport
share rises from \SI{7.0}{\percent} of attributed outages on 27 September to
\SI{85.0}{\percent} on 19 October (a twelve-fold change, at an average
$+2.72$ percentage points per reporting day), and Tennessee moves the same
way over the same window. That is a change in the composition of the outage
large enough that no reasonable reading of the series misses it, and it is the
result; the tests below are a check on it, not the basis for it.

We report them because a monotone recovery makes consecutive days correlated by
construction, so a naive significance test on 23 points would be
anticonservative. Residuals about the linear
trend carry a lag-1 autocorrelation of 0.22, making the 23 days worth roughly
15 independent observations. Correcting for that, and separately by a
moving-block bootstrap that preserves within-block dependence, the weakest of
three $p$-values is $1.5\times10^{-6}$. \ref{app:trend} gives all
three with the autocorrelation diagnostics; we do not lean on them, because with
a single event there is no replication and a $p$-value against a no-trend null
is not what makes this finding useful.

The tail phase contains the 15 October anomaly (\S\ref{sec:res:event}), so it is
fair to ask whether that single day inflates it. It does not: excluding 15
October, the tail is 354 transport site-days against 97 power, a share of
\SI{78.1}{\percent} rather than \SI{80.9}{\percent}.

\paragraph{The inversion is not confined to the mountains}
The obvious objection to a two-state result is that the two states were chosen
after the fact. We therefore ran the same phase decomposition on Georgia and South Carolina,
the two power-dominated states with enough site-days to test.
Table~\ref{tab:phases} gives the result.

\begin{table}[htbp]
\caption{Transport share of attributed outages, by state and event phase. All
four states rise monotonically; what differs is the starting level and how far
the record runs. Georgia and South Carolina leave the reporting area after
7 October, so their restoration phase covers 4--7 October and they have no tail;
Tennessee enters on 28 September, so it has no onset. Damage never exceeds
\SI{4.5}{\percent} of attributed outages in any cell.}
\label{tab:phases}
\small
\setlength{\tabcolsep}{4pt}
\begin{tabular}{lrrrr}
\toprule
      & Onset  & Acute        & Restoration & Tail \\
State & 27 Sep & 28 Sep--3 Oct& 4--10 Oct   & 11--19 Oct \\
\midrule
TN & n/a & \SI{59.8}{\percent} & \SI{84.4}{\percent} & \SI{93.5}{\percent} \\
NC & \SI{7.0}{\percent}  & \SI{47.1}{\percent} & \SI{58.0}{\percent} & \SI{80.9}{\percent} \\
GA & \SI{17.6}{\percent} & \SI{30.6}{\percent} & \SI{54.2}{\percent} & n/a \\
SC & \SI{5.2}{\percent}  & \SI{12.8}{\percent} & \SI{37.5}{\percent} & n/a \\
\bottomrule
\end{tabular}
\end{table}

Every state inverts. Georgia climbs from \SI{17.6}{\percent} to
\SI{54.2}{\percent} and crosses \SI{50}{\percent} before its record ends; South
Carolina, the most power-dominated state in the study, still triples from
\SI{5.2}{\percent} to \SI{37.5}{\percent} in eleven days. The temporal shift
from power to transport is therefore a general property of hurricane recovery,
not a quirk of mountainous terrain.

What terrain changes is the starting level and the endpoint. Tennessee begins
its record already transport-dominated and finishes at \SI{93.5}{\percent};
South Carolina begins at \SI{5.2}{\percent} and, on a record that ends after
eleven days, does not reach parity. This is a more useful result than the
aggregate comparison in \S\ref{sec:res:decomposition}, because it separates two
things the pooled figure conflates: \emph{whether} the mix inverts, which it does
everywhere, and \emph{how far it gets}, which depends on terrain and on how long
the event runs. It also weakens the reading that the coastal states are simply a
different kind of event. They are the same event, observed earlier in its
trajectory and over a shorter window.

The mechanism is unsurprising once stated. Power is restored in days (crews
reconnect feeders, generators are refuelled) while severed fibre must be
physically located, accessed and respliced, often along roads and bridges that
are themselves destroyed. The two causes decay at very different rates, and the
slower one comes to dominate.

\begin{figure}[htbp]
\centering
\includegraphics[width=\linewidth]{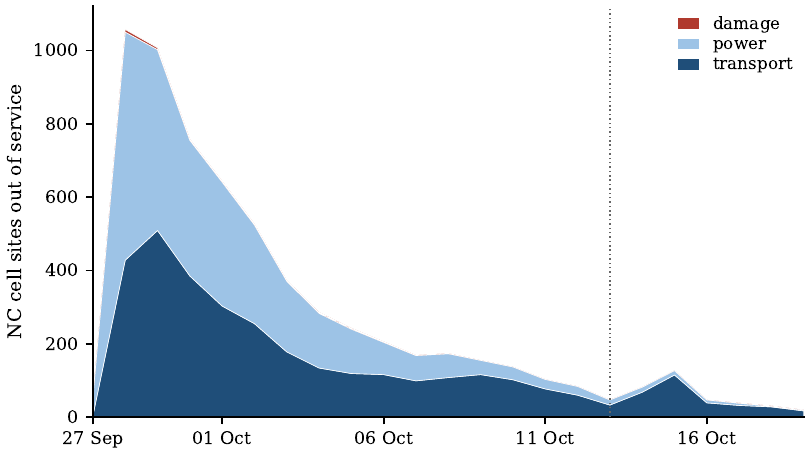}
\caption{North Carolina cell sites out of service by cause. The damage band is
not visible at this scale: it peaks at seven sites on 28 September, against
\num{1079} out that day, and totals 36 site-days across the event
(Table~\ref{tab:decomposition}, \ref{app:daily}). That invisibility is
the finding, not an omission.}
\label{fig:composition}
\end{figure}

Figure~\ref{fig:composition} shows the same data as absolute counts: the total
collapses as power returns, while the transport band persists.

The timing matters for what a mitigation would have to address. Professional
response in the first 48 hours has dedicated satellite and radio channels and
does not depend on the commercial cellular network. Civilian coordination in
weeks two and three does, and that is the window in which the transport share
is highest: \SI{80.9}{\percent} of attributed North Carolina outages in the
tail phase, against \SI{47.1}{\percent} in the acute phase. A mitigation aimed
at the topological failure mode is therefore aimed at the later, longer part of
the outage rather than its peak.

\subsection{Does the rescoping produce the trend?}
\label{sec:res:rescope}

The tail phase straddles a change in the reporting panel: North Carolina's
served count falls from \num{1452} to \num{901} between the 12 and 13 October
reports as 12 counties leave the FCC's disaster area. If the departing counties were
disproportionately power-affected, the apparent rise in transport share would be
an artefact of who left rather than a fact about recovery.

The cause columns are published only at state level, so we cannot compute
county cause shares directly. We can, however, bound the confound: if almost all
of the outage already sat in the counties that were retained, the panel change
can have moved the state share very little. On 12 October, the last day before
the rescope, \textbf{91 of the 96 North Carolina sites out of service
(\SI{94.8}{\percent}) were in the nine counties the panel retained the next
day}. The 12 departing counties held 551 of the \num{1452} sites served
(\SI{37.9}{\percent}) but only \SI{5.2}{\percent} of the outage, and only two of
them (Burke, 3 sites; Caldwell, 2 sites) had any sites out at all. Both figures
are computed on 12 October, where the nine retained counties served exactly the
901 sites that constitute the post-rescope panel.

A panel change that removes \SI{5.2}{\percent} of the outage cannot manufacture
the observed rise from \SI{71.4}{\percent} on 12 October to \SI{82.9}{\percent}
on 14 October, still less the rise from \SI{7.0}{\percent} across the event. We
regard the confound as bounded and small, and note that the trend within the
pre-rescope window alone (27 September to 12 October) is $\rho = 0.85$,
$n = 16$.

\subsection{E3: the 15 October event}
\label{sec:res:event}

North Carolina's outage count reached its post-storm minimum of 54 sites on 13
October, sixteen days after landfall. It then rose on two successive days (to
80 on 14 October and 123 on 15 October) before falling back to 58 on the
16th. Both increases are entirely transport: the transport column goes
$33 \to 68 \to 115$ while power falls $14 \to 14 \to 12$ and damage stays at
zero. The FCC's own 15 October report notes the increase.

We analyse the 14--15 October step in detail, for two reasons. It is the larger
of the two, and it is the one for which an independent instrument shows a
matching excursion (\S\ref{sec:res:ioda}). The 13--14 October step has the same
signature and we have no reason to think it a different event; treating them
separately is conservative, because doing so attributes fewer sites to the
mechanism we identify, not more.

\subsubsection{Reconciling the two numbers}
\label{sec:res:reconcile}

Two figures circulate for this event and they measure different things. Total
sites out rose by 43. Sites out \emph{due to transport} rose by 47. The
difference is not an inconsistency:

\begin{center}
\small
\begin{tabular}{lrrr}
\toprule
             & 14 Oct & 15 Oct & $\Delta$ \\
\midrule
Transport    & 68 & 115 & $+47$ \\
Power        & 14 &  12 & $-2$  \\
Damage       &  0 &   0 & $0$   \\
Unattributed & $-2$ & $-4$ & $-2$ \\
\midrule
Sites out    & 80 & 123 & $+43$ \\
\bottomrule
\end{tabular}
\end{center}

\noindent The four components sum to the change in the total exactly. (The
unattributed residual is negative on these two days because the cause columns
slightly exceed the printed total, a reminder that they are not a strict
partition; see \S\ref{sec:limits}.) Throughout we quote the transport figure,
$+47$, because it is the quantity the analysis is about; the county table below
sums to $+43$ because it counts \emph{sites out}, for which no county-level
cause breakdown is published.

\subsubsection{Eliminating alternatives}

Table~\ref{tab:elimination} tests each candidate explanation against an
independent source.

\begin{table}[htbp]
\caption{Elimination of candidate explanations for the 15 October increase.
Each rests on a primary source; the entire residual is transport.}
\label{tab:elimination}
\small
\begin{tabular}{p{1.45cm}p{4.0cm}l}
\toprule
Candidate & Evidence & Verdict \\
\midrule
Rescoping  & 901 sites served on 13--16 Oct; identical county list
           & excluded \\[2pt]
Weather    & 31 GHCN-Daily stations across all nine counties: max
             0.19\,in on 14 Oct, 0.06\,in on 15
             Oct~\citep{ncei_ghcnd}; 3rd-driest October in NC since
             1895~\citep{ncsco2024october}
           & excluded \\[2pt]
Power      & power-caused outages \emph{fell} 14$\to$12; sites on backup power
             fell 12$\to$10
           & excluded \\[2pt]
Damage     & zero on both days
           & excluded \\[2pt]
Region-wide event
           & cable/wireline subscribers out \emph{fell}
             \num{43696}$\to$\num{40963}; a common regional cause would have
             raised both series (\S\ref{sec:res:ioda})
           & excluded \\
\midrule
Transport  & \textbf{rose 68$\to$115 ($+47$)} & \textbf{residual} \\
\bottomrule
\end{tabular}
\end{table}

The weather exclusion is the one that must carry weight, so we strengthened it.
Rather than rely on a summary account, we retrieved daily precipitation for
every GHCN-Daily station reporting in the nine
counties~\citep{ncei_ghcnd,menne2012ghcnd}: 31
stations, all nine counties represented. Every station recorded
0.00\,in on 13 October. On 14 October the maximum anywhere was
0.19\,in (Hot Springs, Madison County); on 15 October, 0.06\,in
(Burnsville, Yancey County), with every other station at zero or trace. Asheville
Regional Airport recorded 0.03\,in for the entire month.

The row we previously labelled ``broadband'' is relabelled here. Cable
subscriber outages were never a candidate explanation for cell-site outages;
what that series actually rules out is a \emph{region-wide} disruption (a
storm, a regional power event, a metro-scale facility failure) which would
have driven both series in the same direction. It fell while cell sites rose.

\subsubsection{Geographic signature}

We state the limit of this evidence before presenting it. DIRS publishes cause
only at state level, so nothing that follows is an observation of a
county-level cause. What the county
tables give is \emph{sites out}, by county, on two consecutive days; the
attribution of the state-level increase to transport comes from
\S\ref{sec:res:reconcile}, and the step from a spatial pattern to a single
shared facility is an inference from that pattern, not a measurement of it. We
cannot name the facility, the provider, or the sites, and no analysis in this
paper can. \S\ref{sec:res:establishes} states what does and does not follow;
readers who want the bound before the argument should read it now.

\begin{figure}[htbp]
\centering
\includegraphics[width=\linewidth]{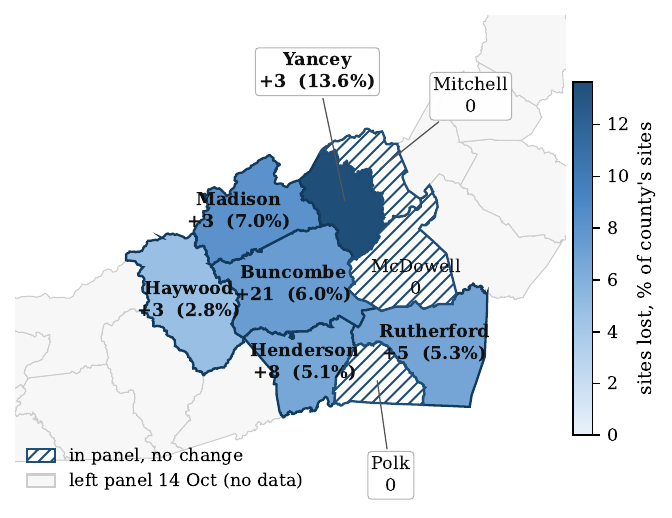}
\caption{Change in cell sites out of service, 14--15 October 2024, across the
nine counties in the reporting panel on both days. Six contiguous counties lose
sites; three do not. The three unaffected counties are not outside the affected
block (McDowell borders Buncombe, Rutherford and Yancey, all affected), which
is why we read the pattern as provider- or route-specific rather than
geographic. Counties in grey left the reporting panel on 14 October and carry no
observation for these days.}
\label{fig:oct15map}
\end{figure}

Table~\ref{tab:oct15} and Figure~\ref{fig:oct15map} give the change per county.

\begin{table}[htbp]
\caption{County-level change, 14--15 October 2024, with exact Poisson
\SI{95}{\percent} intervals on the loss rate and expected losses under a
single per-site rate. ``Nbrs'' counts neighbours \emph{within the nine-county
panel}, not total adjacency. The three counties with no increase are interior
neighbours of affected ones, not peripheral to them.}
\label{tab:oct15}
\small
\setlength{\tabcolsep}{3.6pt}
\begin{tabular}{lrrrrrl}
\toprule
County & Sites & $\Delta$ & \% of county & \SI{95}{\percent} CI & Exp. & Nbrs \\
\midrule
Buncombe   & 348 & $+21$ & \SI{6.0}{\percent}  & 3.7--9.2  & 16.6 & 6 \\
Henderson  & 158 & $+8$  & \SI{5.1}{\percent}  & 2.2--10.0 & 7.5  & 4 \\
Rutherford &  94 & $+5$  & \SI{5.3}{\percent}  & 1.7--12.4 & 4.5  & 4 \\
Madison    &  43 & $+3$  & \SI{7.0}{\percent}  & 1.4--20.4 & 2.1  & 3 \\
Haywood    & 106 & $+3$  & \SI{2.8}{\percent}  & 0.6--8.3  & 5.1  & 3 \\
Yancey     &  22 & $+3$  & \SI{13.6}{\percent} & 2.8--39.9 & 1.1  & 4 \\
\midrule
McDowell   &  83 & $0$   & \SI{0.0}{\percent}  & 0--4.4    & 4.0  & 4 \\
Polk       &  27 & $0$   & \SI{0.0}{\percent}  & 0--13.7   & 1.3  & 2 \\
Mitchell   &  20 & $0$   & \SI{0.0}{\percent}  & 0--18.4   & 1.0  & 2 \\
\midrule
Total      & 901 & $+43$ &                     &           & 43.0 & \\
\bottomrule
\end{tabular}
\end{table}

\paragraph{Among the affected counties, loss is proportional to site count}
The raw rates span a factor of five, from \SI{2.8}{\percent} in Haywood to
\SI{13.6}{\percent} in Yancey, and it would be wrong to call them uniform.
Yancey's figure rests on three sites out of 22 and Madison's on three out of 43;
at those counts the Poisson interval alone spans most of the observed range.
Four counties (Buncombe, Henderson, Rutherford and Madison) do sit close
together, between \SI{5.1}{\percent} and \SI{7.0}{\percent}, but we report that
as a description and attach no test to it: the four were picked out by their
outcome, so testing them for homogeneity would be circular.

What the data support is the weaker statement that among the six affected
counties the loss is \emph{not inconsistent} with being proportional to site
count. A chi-square test of homogeneity against site-count exposure does not
reject a single common per-site rate ($\chi^2 = 4.36$, $\mathrm{df} = 5$,
$p = 0.50$), but at 43 events across six counties that test would also fail to
reject a genuinely varying rate, so it discriminates between the two hypotheses
hardly at all. We report it as a consistency check (the observed spread is
what a single rate would produce, so the spread itself is not evidence
\emph{against} a shared cause), and we place no positive weight on it.

\paragraph{But the panel as a whole is not proportional, and that is the point}
Extending the same test to all nine counties, including the three zeros, gives
$\chi^2 = 12.35$, $\mathrm{df} = 8$, $p = 0.14$ (Monte-Carlo $p = 0.13$ over
$10^6$ multinomial draws, since several expected counts are below five). That
omnibus test does not reject either, and for the same reason carries no
evidential weight: it spends eight degrees of freedom on variation among the
affected counties, which is noise, and dilutes the one contrast that matters.
The statistics that do bear on the question are the two below, which test a
prediction rather than summarise a spread.

The contrast is this. McDowell, Polk and Mitchell hold 130 of the panel's 901
sites (\SI{14.4}{\percent} of the exposure) and took none of the 43
losses, against 6.2 expected. Because those three counties are identified by
their outcome, we do not attach a $p$-value to that comparison directly.
Two versions of it that are \emph{not} conditioned on which counties came up
empty. Neither is decisive on its own (43 events across nine counties is not
much to work with), but both test a geographic prediction rather than the
pattern that suggested it, and both point the same way:

\begin{itemize}
  \item Under a common per-site rate, the counties that happen to record zero
    losses account for \SI{14.4}{\percent} or more of the panel's sites in
    \SI{0.7}{\percent} of $10^6$ simulated draws, and three or more of the
    nine counties come up empty in \SI{6.3}{\percent}.
  \item \textbf{McDowell alone.} McDowell is singled out by geography, not by
    its outcome: it is the only county in the panel interior to the affected
    block, bordering Buncombe, Rutherford and Yancey. It is also the panel's
    fourth largest, with 83 sites and 4.0 expected losses. It recorded zero,
    which a common rate produces with probability $0.016$.
\end{itemize}

\paragraph{The unaffected counties are interior, not peripheral}
This is the part of the pattern that constrains the explanation most. Computing
adjacency directly from Census county boundaries: McDowell borders Buncombe,
Rutherford \emph{and} Yancey, all three of which lost sites; Polk borders
Henderson and Rutherford; Mitchell borders Yancey. None of the three sits
outside the affected block. A purely geographic cause (a storm cell, a
regional power event, a flood) would not skip a county wedged between three
affected neighbours while taking the neighbours on every side. A facility that
is specific to a provider, or to a route, would: it follows the network graph,
not the map. Figure~\ref{fig:oct15map} plots the panel so that the adjacency
claim can be checked directly.

Loss proportional to site count across a contiguous block, with interior
neighbours untouched, is the signature of a \emph{shared upstream path} failing
rather than many independent local failures. If a single provider holds roughly
a quarter to a third of the region's sites and lost about a fifth of its own,
the arithmetic yields the observed \SIrange{5}{7}{\percent} of the total.
Recovery to 58 sites out by the 16 October report (one reporting day, exactly
24 hours between snapshots) is consistent with a fibre splice rather than
reconstruction.

\subsection{E4: external corroboration}
\label{sec:res:ioda}

DIRS is filed by the providers whose networks failed, so the 15 October event
rests, so far, on self-report. We therefore asked whether an independent
instrument saw it.

\begin{figure}[htbp]
\centering
\includegraphics[width=\linewidth]{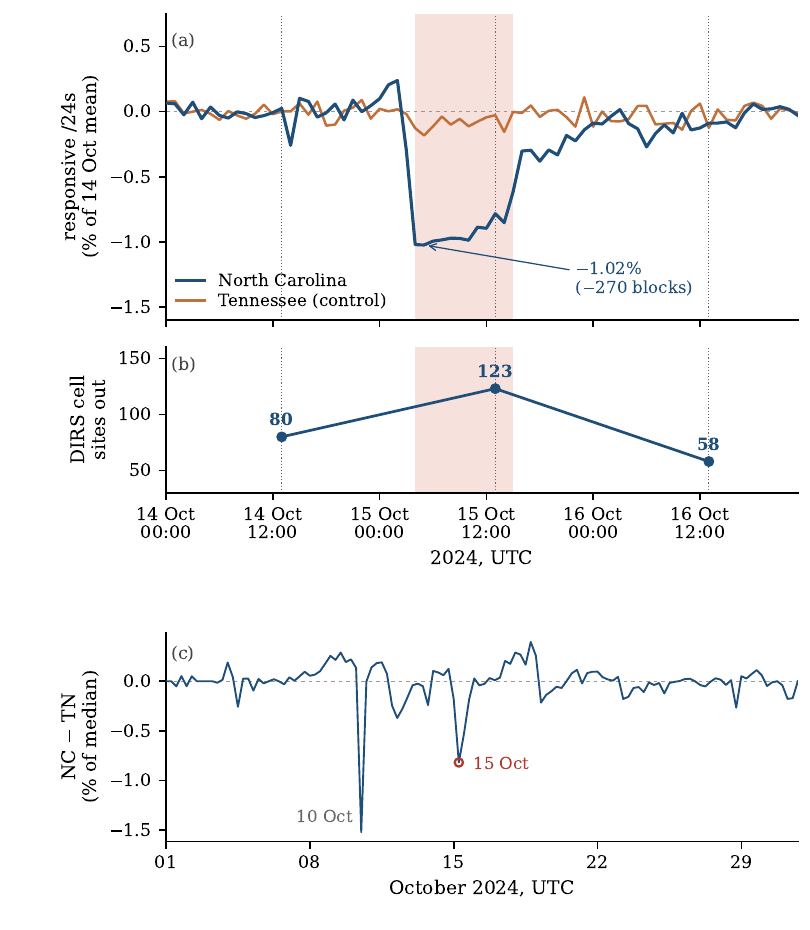}
\caption{(a) IODA active-probe measurement across the event, as a percentage of
each state's own 14 October mean, with Tennessee as a control. (b) The DIRS
series, sampled at the instants the reports describe. Dotted lines mark the
09:00 EDT DIRS snapshots; shading marks the interval in which North Carolina
sits more than \SI{0.5}{\percent} below baseline; the 15 October report was
filed while the dip was in progress. (c) The whole of October, North Carolina
minus the Tennessee control, against a 7-day rolling median. Only two
excursions in the month reach \SI{0.8}{\percent}.}
\label{fig:ioda}
\end{figure}

IODA's \texttt{ping-slash24} signal counts \texttt{/24} address blocks
responding to active probes~\citep{ioda,ioda_api}; the adaptive
\texttt{/24} probing method the signal implements is
that of \citet{quan2013trinocular}. It is a different operator, a
different instrument and a different physical measurement from a provider's
outage filing. Figure~\ref{fig:ioda} plots it for North Carolina and, as a
control, Tennessee (in the same DIRS activation, probed by the same platform,
and reporting no 15 October increase).

\paragraph{Why a cellular backhaul cut is visible to a ping-based signal}
Cell sites do not themselves host responsive \texttt{/24}s, so the connection
needs stating. In this terrain the backhaul carrying a cell site to the network
core and the transport carrying fixed-line subscribers are, very largely, the
same fibre in the same conduit along the same valley route. A cut that strands
cell sites therefore also strands whatever fixed-line address space rides
behind the same facility, and it is the latter that IODA can see. This is not
incidental to the corroboration; it is what makes it meaningful. It is also
why the corroboration is only ever partial: the two populations overlap on the
transport path, not one for one.

North Carolina's responsive-block count is flat at \num{26337} through 14
October (the baseline against which we measure), peaks at \num{26400} at
02:00 UTC on 15 October, and falls to \num{26067} by 05:00 UTC. Relative to
that baseline the drawdown is \SI{1.02}{\percent}, or 270 blocks; measured from
the transient 02:00 peak it would be 333. It remains depressed for twelve hours
and recovers from 15:00 UTC. Tennessee over the same window spans
\SI{0.29}{\percent} end to end and never falls more than \SI{0.18}{\percent}
below its baseline. The dip is the only one in the four-day window and is not
diurnal: the equivalent overnight hours on 14 and 16 October are flat.

\paragraph{How unusual is a \SI{1}{\percent} dip?}
In the raw series, not very; the control is what separates a state-specific
excursion from a platform-wide one. Over the whole of October (124 six-hourly
samples,
Figure~\ref{fig:ioda}c) the \emph{deepest} raw North Carolina drawdown is
\SI{2.5}{\percent} on 11 October, and Tennessee falls \SI{2.6}{\percent} at the
same instant. Excursions that move both states by the same relative amount are
the measurement platform, not the states, and differencing against the control
removes them. In the differenced series the 15 October sample is 4.0 standard
deviations below the October median, and only two of the 124 samples in the
month reach \SI{0.8}{\percent}.

The other is 10 October. North Carolina shows a state-specific excursion of
\SI{1.5}{\percent}
that day with no corresponding movement in DIRS, whose cell-site count fell
monotonically from 9 to 11 October. The two instruments are therefore not in
one-to-one correspondence, and we do not claim they are. What we claim is
narrower: the single day on which DIRS records an
otherwise unexplained increase in cell sites out of service is also one of only
two days in the month on which IODA records a North-Carolina-specific
connectivity excursion, and the two coincide to within hours.

The routing plane says the failure was not upstream. IODA's BGP signal for
North Carolina moves by 3 prefixes out of \num{79007} (\SI{0.004}{\percent})
across the entire twelve-hour window, having last stepped a day earlier. A
withdrawn transit path or a failed core facility would appear here; an
access-layer or backhaul failure, in which the prefixes remain advertised while
the equipment behind them becomes unreachable, would not. That is the shape of
what we observe.

The timing is the part worth stating carefully. DIRS reports describe
09:00 EDT (13:00 UTC) on their date. At that instant on 14, 15
and 16 October, IODA's count reads \num{26344}, \num{26131} and \num{26314}:
down 213 blocks at the 15 October snapshot, recovered by the 16th. The two
instruments, measuring different things by different means, move together and
recover together. IODA also shows recovery beginning about two hours after the
15 October snapshot, which is consistent with the single-reporting-day
restoration DIRS records.

\paragraph{Reconciling this with the ``region-wide event'' exclusion}
Table~\ref{tab:elimination} excludes a region-wide cause partly because
cable and wireline subscribers out \emph{fell} on 15 October, from \num{43696}
to \num{40963}. Here we read a \emph{fall} in a fixed-line connectivity measure
as evidence that something happened. Both readings are correct, but the
apparent tension needs resolving explicitly.

The two are compatible for three reasons, in increasing order of importance.
The series measure different populations: DIRS wireline counts subscribers of
the cable and wireline providers that chose to file, while IODA counts
responsive address space across every network in the state, including transport
providers and networks that file nothing. They have different resolutions: the
DIRS figure is one number per day, moving against a restoration trend that was
returning thousands of subscribers daily, so a twelve-hour excursion of a few
hundred blocks is not separable from it. And the two rows are not making the
same claim. The Table~\ref{tab:elimination}
row does not assert that no fixed-line disruption occurred on 15 October; it
rules out a \emph{region-wide} cause, which would have driven the subscriber
series sharply upward by thousands rather than leaving it falling. A localised
transport facility failing, affecting one provider's backhaul and whatever
address space shares its path, is exactly the kind of event that shows up as a
\SI{1}{\percent} probing excursion while a statewide net subscriber count
continues to recover. Both observations are consistent with the same
explanation.

\paragraph{Limits of the corroboration}
IODA publishes no county-level series. Entity codes for all nine counties
resolve, but every datasource returns an empty series, and IODA's own
documentation describes country, region and network granularity only. This
corroboration is therefore statewide: it confirms that something removed
connectivity in North Carolina, in the right direction, on the right day, for
about the right duration, and that Tennessee did not experience it. It cannot
distinguish the six affected counties from the three unaffected ones. That
discrimination remains available only from DIRS, or from NORS.

\subsubsection{What the 15 October analysis establishes}
\label{sec:res:establishes}

Eliminating rescoping, weather, power, damage and a region-wide event is well
supported: each rests on a primary source, and the timing is now independently
corroborated. The inference to a \emph{single shared} backhaul facility remains
indirect: we cannot name it, because DIRS does not identify providers or
sites, and IODA cannot localise below state level.

What the event does establish does not depend on that mechanism. On 15 October
there was no rainfall, no rise in power-caused outages and no reported damage;
the 47 sites were, by the filings' own accounting, intact and powered. A change
in \emph{connectivity alone} removed service across six counties within a
single reporting interval, 17 days after landfall: a failure mode distinct
from physical destruction, and observable in a public filing.

\subsection{Implications}

Taken together the four analyses say something narrower than ``backhaul is the
problem'', and more useful. Cell sites overwhelmingly survived Helene physically
(\S\ref{sec:res:decomposition}). Whatever mix of power and transport took them
out of service, that mix shifted toward transport in every state we can measure,
and kept shifting for as long as each record runs (\S\ref{sec:res:temporal}).
And a transport failure alone can remove service from a whole region in the
third week of a recovery, with no weather, no power loss and no damage
(\S\ref{sec:res:event}, \S\ref{sec:res:ioda}). Three consequences follow.

\textbf{Resilience investment is mistargeted where terrain constrains routes.}
Backup power is the standard hardening measure for cell sites and it addresses
\SI{47.3}{\percent} of this event in North Carolina and \SI{63.2}{\percent} of
it pooled across all six states. In the
mountains, transport diversity addresses the larger and more persistent share,
and it is not systematically reported, mandated or measured. The remedy is not
hypothetical: a regional broadcaster stayed on the air through the event by
substituting satellite connectivity for its terrestrial
backhaul~\citep{connelly2025radio}: the same substitution, one layer over,
made by an operator who could choose it. The FCC's 2026
harmonisation of transport-facility reporting is a necessary precondition for
even observing it~\citep{fcc2026resilient}.

\textbf{Recovery work may cause re-failure.} The 15 October event occurred amid
intense reconstruction. Debris removal and rebuilding involve heavy earthmoving,
which severs fibre. If this generalises, the period of maximum reconstruction
activity is also a period of elevated topological failure, a feedback loop
that current planning does not represent.

\textbf{Application architecture assumes the wrong failure.} Every mainstream
communication application requires a path to a distant server, so a region whose
local infrastructure survives but whose backhaul is severed loses all of them at
once.

%% ===========================================================================
\section{Limitations and Future Work}
\label{sec:limits}

\subsection{Limitations of the source}

\textbf{The cause columns are not a strict partition.} The three counts do not
sum to the printed total: they fall short by \SI{6.9}{\percent} of North
Carolina site-days overall and \SI{7.6}{\percent} pooled, and on a few days
slightly exceed it. We report both denominators and the residual wherever a
headline figure appears (\S\ref{sec:definitions}), and test both conclusions
against the residual adversarially in \S\ref{sec:res:decomposition}. The
outcome is asymmetric and worth restating here: the damage share is a floor
(ceiling \SI{8.6}{\percent} pooled if the entire residual were damage), while
the North Carolina transport-over-power ordering would reverse if more than
\SI{66}{\percent} of that state's residual were really power. We find no
evidence of such a skew, but we cannot exclude it, and a reader who assumes it
should read the North Carolina ordering as unresolved and rely on Tennessee,
whose residual is \SI{0.8}{\percent}.

\textbf{``Damage'' means damage to the cell site.} It does not mean the network
suffered no physical destruction; Helene destroyed more than \num{1700} miles of
fibre. Much of what DIRS records as a transport outage is downstream of physical
damage to a transport facility. Our contribution is to locate that damage on the
network graph, not to deny it.

\textbf{The reporting area is rescoped mid-event.} Denominators change
discontinuously: North Carolina falls from \num{1452} to \num{901} sites
between the 12 and 13 October reports. A naive reading of the raw subscriber series would
show an overnight fall from \num{325348} to \num{84085} on 8 October; this is
Georgia and South Carolina leaving the reporting area, not recovery. We use
within-state shares throughout and bound the residual confound in
\S\ref{sec:res:rescope}.

\textbf{Reporting is voluntary and unattributed.} DIRS does not identify which
provider filed what, nor which specific site failed. We characterise causes in
aggregate but cannot attribute an outage to an operator or a facility.

\textbf{Cause is self-reported.} ``Transport'' is the filing provider's own
classification; we have no means to audit it, and a site simultaneously lacking
power and backhaul could plausibly be filed under either. The 15 October event
is the one place we obtain external corroboration (\S\ref{sec:res:ioda}), and
only at state granularity.

\subsection{Limitations of the analysis}

\textbf{Cell sites are not people.} A site-weighted share is not a
population-weighted one: a rural site serves far fewer subscribers than an urban
one. Our findings characterise infrastructure, not experienced outage.
Population-weighting county outage rates for western North Carolina on 29
September yields roughly \num{350000} residents in counties experiencing
transport-caused outage, but we regard this as order-of-magnitude only.

\textbf{One event, and a partly terrain-dependent result.} These are the causes
of one hurricane. Within it, transport \emph{dominance} over the full record
holds in two mountainous states and not in four others, while the temporal
shift toward transport holds in all four states long enough to test
(\S\ref{sec:res:temporal}). We scope each claim accordingly, but cannot test the
terrain mechanism itself with six
states and no route-diversity data.

\textbf{The trend is descriptive.} With one event there is no replication. We
report autocorrelation-corrected significance (\S\ref{sec:res:temporal}) because
reviewers of measurement work reasonably ask for it, not because a $p$-value
carries inferential weight here.

\textbf{This is an observational study, not a natural experiment.} The 15
October event involves no as-if-random assignment. It is an isolating event: an
anomaly whose alternative explanations can be eliminated against primary
sources. We think that is respectable, and we do not claim more for it.

\textbf{A transport-caused outage is not automatically recoverable.} A site
without backhaul cannot serve traffic either, since the mobile core is upstream.
Our decomposition measures how much of the failure was \emph{topological rather
than physical}, an upper bound on what architectural mitigation could
address, not a quantity of service that was available and unused.

\subsection{Future work}

\textbf{Other DIRS activations.} The Commission activates DIRS for every major
disaster, and each activation produces a comparable record. Applying this
method across events is the direct test of the terrain hypothesis: it would
separate whether transport dominance is a property of mountainous terrain, of
Helene specifically, or of modern cellular infrastructure generally. Our
six-state comparison suggests the first, and cannot establish it.

\textbf{Confirming the 15 October mechanism.} The corresponding NORS filings
would name the provider, the facility and the root cause. NORS is confidential
under 47~C.F.R.~\S4.2, but the Commission processes requests for confidential
records under \S0.461, and an aggregated disclosure across four or more
providers would suffice to confirm or refute the shared-path hypothesis without
exposing competitively sensitive detail.

\textbf{Sub-state external measurement.} IODA corroborates the 15 October event
at state level (\S\ref{sec:res:ioda}) but publishes nothing finer. Probing data
with county or ASN resolution (or IODA's per-network signals matched to the
providers serving western North Carolina) would test the county-level pattern
directly, and more generally allow the cause-labelled DIRS record to be
cross-validated against externally observed outages.

\textbf{Architectural implications.} A failure mode in which local
infrastructure survives but the path out does not is, in principle, addressable
by systems that do not require a path out. Evaluating partition-tolerant designs
against this trace, rather than against synthetic outage models, is the
direction we intend to pursue.

%% ===========================================================================
\section{Conclusions}

The largest cell-site outage in the FCC's public disaster record was not, in
the main, caused by destroyed cell sites. Across six states, damage at the
site accounts for \SI{1.1}{\percent} of attributed cell-site-days and never
exceeds \SI{3.8}{\percent} anywhere. The sites were standing.

What took them out divides by terrain. Pooled across the whole event, commercial
power remains the larger cause. But in the mountains of western North Carolina,
where Helene's outage was largest and longest, severed backhaul accounts for
\SI{52.2}{\percent} of attributed cell-site-days against \SI{47.3}{\percent} for
power, and its share grows across the event from \SI{7.0}{\percent} to
\SI{85.0}{\percent} until it accounts for essentially all remaining outages, a
pattern Tennessee independently reproduces. On 15 October, in the third week of
recovery, 47 sites across six contiguous counties lost their backhaul with no
weather, no power loss and no damage of any kind, and an independent probing
platform recorded the same twelve hours.

The towers were, for the most part, standing and powered. What they had lost was
the path to the network core.

%% ===========================================================================
%% =========================================================================
%% Elsevier back matter. All three statements below are required by the
%% editorial system; the journal will ask for them separately if they are not
%% in the manuscript.
%%
%% The artifact goes up with the submission as supplementary material. Add the
%% repository DOI to the statement below once the deposit exists; Elsevier's
%% own Mendeley Data is free and issues one, and Editorial Manager links it to
%% the article.
%% =========================================================================
\section*{CRediT authorship contribution statement}

%% =========================================================================
%% CONFIRM WITH EACH CO-AUTHOR BEFORE SUBMISSION. CRediT roles are a factual
%% claim about who did what, and Elsevier asks the corresponding author to
%% attest to them.
%%
%% Addeh and Konan are credited with the two writing roles, per Ola. CRediT
%% separates them: "original draft" is composing the text, "review and editing"
%% is critical revision of someone else's. If their contribution was revising
%% rather than drafting, delete "Writing (original draft)" from both.
%% =========================================================================
\textbf{Oluseyi Olukola:} Conceptualization, Methodology, Software, Formal
analysis, Data curation, Investigation, Validation, Visualization, Writing (original draft), Writing (review and editing).
\textbf{Oare Danielle Addeh:} Writing (original draft), Writing (review and editing).
\textbf{Esther Abiodun Konan:} Writing (original draft), Writing (review and editing).
\textbf{Nick Rahimi:} Conceptualization, Methodology, Supervision, Validation,
Writing (review and editing), Project administration.

\section*{Declaration of competing interest}

The authors declare that they have no known competing financial interests or
personal relationships that could have appeared to influence the work reported
in this paper.

\section*{Data availability}

The reconstructed dataset (\num{80} state-days and \num{580} county-days with
per-cause breakdown, the extraction and reconciliation code, the statistical
checks reported in \S\ref{sec:results}, and the figure sources) accompanies
this submission as supplementary material, and will be deposited in a public
repository under a CC-BY-4.0 licence, with a citable DOI, on acceptance. All
source documents are public FCC filings, individually cited by document number; the
NCEI and IODA queries are given as complete request URLs so that both external
checks can be re-run exactly.

\bibliographystyle{elsarticle-harv}
\bibliography{references}

%% ===========================================================================
\appendix

\section{Full daily record, North Carolina}
\label{app:daily}

Table~\ref{tab:daily} gives the complete per-day North Carolina record
underlying Figures~\ref{fig:inversion} and~\ref{fig:composition}. Denominator
changes are marked and no values are interpolated. The panel narrows between
12 and 13 October, which is where the served count falls from \num{1452} to
901.

\begin{table}[htbp]
\caption{North Carolina daily record. ``Share'' is the transport share,
computed against the sum of the three cause columns (\S\ref{sec:definitions}).
Note the served-count changes on 29 September and 13 October, and the two-day
rise on 14--15 October analysed in \S\ref{sec:res:event}.}
\label{tab:daily}
\small
\begin{tabular}{lrrrrrr}
\toprule
Date & Served & Out & Dmg & Trans & Power & Share \\
\midrule
27 Sep & \num{1452} & 58   & 1 & 4   & 52  & \SI{7.0}{\percent}  \\
28 Sep & \num{1452} & \num{1079} & 7 & 428 & 622 & \SI{40.5}{\percent} \\
29 Sep & \num{1556} & \num{1034} & 5 & 509 & 493 & \SI{50.5}{\percent} \\
30 Sep & \num{1452} & 784  & 3 & 385 & 370 & \SI{50.8}{\percent} \\
1 Oct  & \num{1452} & 707  & 3 & 303 & 338 & \SI{47.0}{\percent} \\
2 Oct  & \num{1448} & 554  & 3 & 255 & 269 & \SI{48.4}{\percent} \\
3 Oct  & \num{1448} & 410  & 2 & 178 & 193 & \SI{47.7}{\percent} \\
4 Oct  & \num{1448} & 321  & 3 & 134 & 149 & \SI{46.9}{\percent} \\
5 Oct  & \num{1448} & 277  & 3 & 119 & 121 & \SI{49.0}{\percent} \\
6 Oct  & \num{1448} & 244  & 2 & 116 & 88  & \SI{56.3}{\percent} \\
7 Oct  & \num{1448} & 209  & 1 & 99  & 69  & \SI{58.6}{\percent} \\
8 Oct  & \num{1448} & 213  & 1 & 108 & 65  & \SI{62.1}{\percent} \\
9 Oct  & \num{1448} & 184  & 0 & 116 & 39  & \SI{74.8}{\percent} \\
10 Oct & \num{1448} & 154  & 0 & 102 & 35  & \SI{74.5}{\percent} \\
11 Oct & \num{1448} & 115  & 0 & 77  & 26  & \SI{74.8}{\percent} \\
12 Oct & \num{1452} & 96   & 0 & 60  & 24  & \SI{71.4}{\percent} \\
13 Oct & 901        & 54   & 0 & 33  & 14  & \SI{70.2}{\percent} \\
14 Oct & 901        & 80   & 0 & 68  & 14  & \SI{82.9}{\percent} \\
15 Oct & 901        & 123  & 0 & 115 & 12  & \SI{90.6}{\percent} \\
16 Oct & 901        & 58   & 0 & 39  & 8   & \SI{83.0}{\percent} \\
17 Oct & 901        & 39   & 1 & 32  & 6   & \SI{82.1}{\percent} \\
18 Oct & 901        & 32   & 1 & 28  & 2   & \SI{90.3}{\percent} \\
19 Oct & 901        & 20   & 0 & 17  & 3   & \SI{85.0}{\percent} \\
\bottomrule
\end{tabular}
\end{table}

\section{Trend statistics}
\label{app:trend}

Table~\ref{tab:trend} gives the trend statistics summarised in
\S\ref{sec:res:temporal}. They are reported here rather than in the body
because the finding is the size of the shift, not the rejection of a no-trend
null; these numbers establish only that serial correlation does not account for
the shift.

\begin{table}[htbp]
\caption{Trend in the North Carolina daily transport share. The naive
$p$-value is anticonservative because a monotone recovery makes consecutive
days correlated by construction; the two corrections address that. All three
are reported so the correction can be judged.}
\label{tab:trend}
\small
\begin{tabular}{lr}
\toprule
Statistic & Value \\
\midrule
Observations $n$                        & 23 \\
Spearman $\rho$                         & 0.925 \\
Pearson $r$                             & 0.919 \\
OLS slope                               & $+2.72$ pp/day \\
\midrule
Residual lag-1, $\mathrm{corr}(e_t, e_{t-1})$ & 0.215 \\
Residual lag-1, Yule--Walker            & 0.154 \\
Durbin--Watson                          & 1.16 \\
Effective $n$ (Bartlett, on 0.215)      & 14.9 \\
\midrule
$p$, naive                              & $2.8\times10^{-10}$ \\
$p$, autocorrelation-adjusted           & $1.5\times10^{-6}$ \\
$p$, moving-block bootstrap             & $<10^{-4}$ \\
\bottomrule
\end{tabular}
\end{table}

Two details are worth stating so they are not mistaken for errors. First, the
Durbin--Watson statistic (1.16) is lower than the $2(1-\hat\rho)$ approximation
would suggest from either lag-1 estimate (1.57 and 1.69). The gap is an end
effect, not a contradiction: the exact identity is
$\mathrm{DW} = 2 - 2\hat\rho_{\mathrm{YW}} - (e_1^2 + e_n^2)/\sum e_t^2$, and
here the two endpoint residuals carry \SI{53}{\percent} of the residual sum of
squares, almost all of it the 27 September onset day, which sits 26 points
below the fitted line. At $n=23$ that correction term is not negligible, as it
is asymptotically. Second, we use the larger of the two lag-1 estimates for the
Bartlett correction, which is the conservative choice: the Yule--Walker
estimate gives an effective $n$ of 16.9 and a smaller $p$ of
$2.2\times10^{-7}$. The moving-block bootstrap uses $10^4$ resamples of
contiguous 4-day blocks, the timescale of the recovery itself.

\section{External check queries}
\label{app:queries}

Both external checks are reproducible from public APIs. Precipitation
(\S\ref{sec:res:event}) uses the NCEI Access Data Service, dataset
\texttt{daily-summaries}, element \texttt{PRCP}, for 31 stations across the nine
counties. Active-probe corroboration (\S\ref{sec:res:ioda}) uses the IODA v2
API, signal \texttt{ping-slash24}, region entity 4444 (North Carolina) and 4446
(Tennessee), over \texttt{from=1728864000} to \texttt{until=1729123200}. Full
request URLs and the retrieved series are included in the artifact.

\clearpage

\end{document}

%% file: figures/methodology.tex
% Methodology pipeline. Kept in its own file so it can be edited without
% touching the manuscript, and \input{} into main.tex.
%
% The visual argument is the FORK: two extractions run independently from the
% same PDFs and are only compared afterwards. That is what makes the
% reconciliation a genuine check rather than a self-consistency test.
\begin{tikzpicture}[
    font=\footnotesize,
    node distance=0pt,
    stage/.style={draw=black!55, rounded corners=2pt, align=center,
                  inner sep=4pt, minimum height=12mm, fill=white, line width=0.5pt},
    src/.style={stage, fill=black!5, text width=17mm},
    ext/.style={stage, text width=21mm},
    rec/.style={stage, fill=blue!7, draw=blue!55, text width=21mm},
    ana/.style={stage, fill=orange!8, draw=orange!65, text width=23mm},
    lbl/.style={font=\scriptsize\itshape, text=black!65},
    flow/.style={-{Stealth[length=4pt]}, draw=black!60, line width=0.5pt},
]

% ---- stage 1: acquisition ------------------------------------------------
\node[src] (pdfs) {\textbf{FCC DIRS}\\24 daily reports\\26 Sep--19 Oct};

% ---- stage 2: two independent extractions --------------------------------
\node[ext, right=11mm of pdfs, yshift=9mm] (extA)
     {Extraction A\\\scriptsize state totals\\\scriptsize + cause columns};
\node[ext, right=11mm of pdfs, yshift=-9mm] (extB)
     {Extraction B\\\scriptsize county totals\\\scriptsize per state};

% ---- stage 3: reconciliation --------------------------------------------
\node[rec, right=11mm of pdfs.east, xshift=32mm, text width=23mm] (rec)
     {\textbf{Reconcile}\\[1pt]\scriptsize$\sum$\,county $=$ state\\\scriptsize 45/45 days exact};

% ---- dataset -------------------------------------------------------------
\node[stage, right=9mm of rec, text width=20mm, fill=black!5] (data)
     {\textbf{Dataset}\\\scriptsize 80 state-days\\\scriptsize 580 county-days};

% ---- stage 4: analyses ---------------------------------------------------
\node[ana, right=10mm of data, yshift=13mm] (e1)
     {\textbf{E1} decomposition\\\scriptsize cause shares};
\node[ana, right=10mm of data] (e2)
     {\textbf{E2} temporal\\\scriptsize trend, phases};
\node[ana, right=10mm of data, yshift=-13mm] (e3)
     {\textbf{E3} 15 Oct event\\\scriptsize alternative elimination};

% ---- edges ---------------------------------------------------------------
\draw[flow] (pdfs.east) -- ++(4mm,0) |- (extA.west);
\draw[flow] (pdfs.east) -- ++(4mm,0) |- (extB.west);
\draw[flow] (extA.east) -| ($(rec.north)+(0,4mm)$) -- (rec.north);
\draw[flow] (extB.east) -| ($(rec.south)-(0,4mm)$) -- (rec.south);
\draw[flow] (rec.east) -- (data.west);
\draw[flow] (data.east) -- ++(4mm,0) |- (e1.west);
\draw[flow] (data.east) -- ++(4mm,0) |- (e2.west);
\draw[flow] (data.east) -- ++(4mm,0) |- (e3.west);

% ---- annotations ---------------------------------------------------------
% The two extraction boxes now nearly touch, so this label no longer
% fits between them; it goes in the gap after the fork instead.
\node[lbl, rotate=90] at ($(pdfs.east)+(7.5mm,0)$) {independent};
\node[lbl, below=6mm of rec, text width=26mm, align=center]
     {cross-validation,\\not self-consistency};

\end{tikzpicture}

%% file: references.bib
@inproceedings{dainotti2011outages,
  author    = {Dainotti, Alberto and Squarcella, Claudio and Aben, Emile and
               Claffy, Kimberly C. and Chiesa, Marco and Russo, Michele and
               Pescap{\'e}, Antonio},
  title     = {Analysis of Country-Wide Internet Outages Caused by Censorship},
  booktitle = {Proceedings of the 2011 ACM SIGCOMM Conference on Internet
               Measurement Conference (IMC '11)},
  pages     = {1--18},
  address   = {Berlin, Germany},
  month     = nov,
  year      = {2011},
  publisher = {Association for Computing Machinery},
  doi       = {10.1145/2068816.2068818}
}

@article{dainotti2014outages,
  author  = {Dainotti, Alberto and Squarcella, Claudio and Aben, Emile and
             Claffy, Kimberly C. and Chiesa, Marco and Russo, Michele and
             Pescap{\'e}, Antonio},
  title   = {Analysis of Country-Wide Internet Outages Caused by Censorship},
  journal = {IEEE/ACM Transactions on Networking},
  volume  = {22},
  number  = {6},
  pages   = {1964--1977},
  month   = dec,
  year    = {2014},
  issn    = {1063-6692},
  doi     = {10.1109/TNET.2013.2291244}
}

@inproceedings{bischof2023destination,
  author    = {Bischof, Zachary S. and Pitcher, Kennedy and Carisimo, Esteban and
               Meng, Amanda and Nunes, Rafael Bezerra and
               Padmanabhan, Ramakrishna and Roberts, Margaret E. and
               Snoeren, Alex C. and Dainotti, Alberto},
  title     = {Destination Unreachable: Characterizing Internet Outages and
               Shutdowns},
  booktitle = {Proceedings of the ACM SIGCOMM 2023 Conference},
  pages     = {608--621},
  address   = {New York, NY, USA},
  month     = sep,
  year      = {2023},
  publisher = {Association for Computing Machinery},
  doi       = {10.1145/3603269.3604883}
}

@misc{ioda,
  author       = {{Georgia Institute of Technology Internet Intelligence Lab}},
  title        = {{IODA}: Internet Outage Detection and Analysis},
  howpublished = {Online platform, \url{https://ioda.inetintel.cc.gatech.edu/}},
  year         = {2026}
}

@article{quan2013trinocular,
  author  = {Quan, Lin and Heidemann, John and Pradkin, Yuri},
  title   = {Trinocular: Understanding Internet Reliability through Adaptive
             Probing},
  journal = {ACM SIGCOMM Computer Communication Review},
  volume  = {43},
  number  = {4},
  pages   = {255--266},
  month   = aug,
  year    = {2013},
  issn    = {0146-4833},
  doi     = {10.1145/2534169.2486017},
  note    = {Proceedings of ACM SIGCOMM 2013}
}

@techreport{heidemann2012sandy,
  author      = {Heidemann, John and Quan, Lin and Pradkin, Yuri},
  title       = {A Preliminary Analysis of Network Outages During Hurricane Sandy},
  institution = {USC/Information Sciences Institute},
  number      = {ISI-TR-2008-685b},
  month       = nov,
  year        = {2012},
  note        = {Correction February 2013},
  url         = {https://ant.isi.edu/~johnh/PAPERS/Heidemann12d.html}
}

@inproceedings{cho2011japan,
  author    = {Cho, Kenjiro and Pelsser, Cristel and Bush, Randy and Won, Youngjoon},
  title     = {The {Japan} Earthquake: The Impact on Traffic and Routing Observed
               by a Local {ISP}},
  booktitle = {Proceedings of the Special Workshop on Internet and Disasters
               (SWID '11), co-located with ACM CoNEXT 2011},
  articleno = {2},
  pages     = {1--8},
  address   = {Tokyo, Japan},
  year      = {2011},
  publisher = {Association for Computing Machinery},
  doi       = {10.1145/2079360.2079362}
}

@inproceedings{padmanabhan2019weather,
  author    = {Padmanabhan, Ramakrishna and Schulman, Aaron and Levin, Dave and
               Spring, Neil},
  title     = {Residential Links under the Weather},
  booktitle = {Proceedings of the ACM Special Interest Group on Data
               Communication (SIGCOMM '19)},
  pages     = {145--158},
  address   = {Beijing, China},
  month     = aug,
  year      = {2019},
  publisher = {Association for Computing Machinery},
  doi       = {10.1145/3341302.3342084}
}

@inproceedings{durairajan2018lightsout,
  author    = {Durairajan, Ramakrishnan and Barford, Carol and Barford, Paul},
  title     = {Lights Out: Climate Change Risk to Internet Infrastructure},
  booktitle = {Proceedings of the Applied Networking Research Workshop (ANRW '18)},
  pages     = {9--15},
  address   = {Montreal, QC, Canada},
  month     = jul,
  year      = {2018},
  publisher = {Association for Computing Machinery},
  doi       = {10.1145/3232755.3232775}
}

@inproceedings{maitland2018puertorico,
  author    = {Maitland, Carleen and Peha, Jon M.},
  title     = {Wireless Network Recovery Following Natural Disaster: {Puerto
               Rico} after {Hurricane Maria}},
  booktitle = {TPRC 46: The 46th Research Conference on Communication,
               Information and Internet Policy},
  year      = {2018},
  note      = {SSRN Abstract ID 3142393, posted 19 March 2018},
  url       = {https://ssrn.com/abstract=3142393}
}

@article{booker2010cellular,
  author  = {Booker, Graham and Torres, Jacob and Guikema, Seth and
             Sprintson, Alex and Brumbelow, Kelly},
  title   = {Estimating Cellular Network Performance during Hurricanes},
  journal = {Reliability Engineering \& System Safety},
  volume  = {95},
  number  = {4},
  pages   = {337--344},
  year    = {2010},
  issn    = {0951-8320},
  doi     = {10.1016/j.ress.2009.11.003}
}

@article{du2024telecomoutage,
  author  = {Du, Ao},
  title   = {Data-Driven Telecommunication Outage Prediction during Hurricane
             Events},
  journal = {ASCE-ASME Journal of Risk and Uncertainty in Engineering Systems,
             Part A: Civil Engineering},
  volume  = {10},
  number  = {3},
  pages   = {04024046},
  year    = {2024},
  issn    = {2376-7642},
  doi     = {10.1061/AJRUA6.RUENG-1285},
  note    = {Uses county-level telecommunication outage data from ten recent
             US hurricanes}
}

@inproceedings{griffith2015resiliency,
  author    = {Griffith, David and Rouil, Richard and Izquierdo, Antonio
               and Golmie, Nada},
  title     = {Measuring the Resiliency of Cellular Base Station Deployments},
  booktitle = {2015 IEEE Wireless Communications and Networking Conference
               (WCNC)},
  pages     = {1625--1630},
  address   = {New Orleans, LA, USA},
  year      = {2015},
  publisher = {IEEE},
  doi       = {10.1109/WCNC.2015.7127711}
}

@inproceedings{malandrino2017disasters,
  author    = {Malandrino, Francesco and Chiasserini, Carla Fabiana},
  title     = {Quantifying and Minimizing the Impact of Disasters on Wireless
               Communications},
  booktitle = {Proceedings of the First CoNEXT Workshop on ICT Tools for
               Emergency Networks and DisastEr Relief (WICTEND '17)},
  pages     = {26--30},
  address   = {Incheon, Republic of Korea},
  month     = dec,
  year      = {2017},
  publisher = {Association for Computing Machinery},
  doi       = {10.1145/3152896.3152902}
}

@article{yang2017backhaul,
  author  = {Yang, Sen and He, Yan and Ge, Zihui and Wang, Dongmei and Xu, Jun},
  title   = {Predictive Impact Analysis for Designing a Resilient Cellular
             Backhaul Network},
  journal = {Proceedings of the ACM on Measurement and Analysis of Computing
             Systems},
  volume  = {1},
  number  = {2},
  articleno = {30},
  pages   = {1--33},
  month   = dec,
  year    = {2017},
  issn    = {2476-1249},
  doi     = {10.1145/3154488}
}

@article{kuhn1997pstn,
  author  = {Kuhn, D. Richard},
  title   = {Sources of Failure in the Public Switched Telephone Network},
  journal = {Computer},
  volume  = {30},
  number  = {4},
  pages   = {31--36},
  month   = apr,
  year    = {1997},
  issn    = {0018-9162},
  doi     = {10.1109/2.585151}
}

@techreport{gallagher2024helene,
  author      = {Gallagher, Jill C.},
  title       = {Restoration of Cell Phone Services: {Hurricane Helene}},
  institution = {Congressional Research Service},
  type        = {CRS In Focus},
  number      = {IF12779},
  month       = oct # {~8},
  year        = {2024},
  url         = {https://www.congress.gov/crs-product/IF12779}
}

@techreport{crs2025cellularoutage,
  author       = {Congressional Research Service},
  title       = {Cellular Network Outage Reporting and Restoration During
                 Disasters},
  institution = {Congressional Research Service},
  type        = {CRS Report},
  number      = {R48776},
  month       = dec # {~22},
  year        = {2025},
  url         = {https://www.congress.gov/crs-product/R48776},
  note        = {Describes NORS, DIRS and the Mandatory Disaster Response
                 Initiative and their data limitations}
}

@misc{fcc2026resilient,
  author       = {{Federal Communications Commission}},
  title        = {Resilient Networks; Amendments to Part 4 of the Commission's
                  Rules Concerning Disruptions to Communications; New Part 4 of
                  the Commission's Rules Concerning Disruptions to
                  Communications},
  howpublished = {Third Report and Order, PS Docket Nos.\ 21-346 and 15-80,
                  ET Docket No.\ 04-35, FCC 26-34},
  month        = may,
  year         = {2026},
  note         = {Adopted 20 May 2026; released 21 May 2026},
  url          = {https://docs.fcc.gov/public/attachments/FCC-26-34A1.pdf}
}

@misc{fcc2026resilientfr,
  author       = {{Federal Communications Commission}},
  title        = {Resilient Networks; Concerning Disruptions to Communications},
  howpublished = {Final rule, 91 Fed.\ Reg.\ 39516 (June 30, 2026)
                  (to be codified at 47 C.F.R.\ pt.\ 4)},
  month        = jun,
  year         = {2026},
  note         = {FR Doc.\ 2026-13155; effective 30 June 2026 except
                  47 C.F.R.\ \S~4.18},
  url          = {https://www.federalregister.gov/documents/2026/06/30/2026-13155}
}

@misc{cfr47part4,
  author       = {Code of Federal Regulations},
  title        = {Disruptions to Communications},
  howpublished = {47 C.F.R.\ pt.\ 4 (2026)},
  year         = {2026},
  url          = {https://www.ecfr.gov/current/title-47/chapter-I/subchapter-A/part-4}
}

@misc{fcc_dirs_helene,
  author       = {{Federal Communications Commission}},
  title        = {Communications Status Reports for Areas Impacted by
                  Hurricane Helene},
  howpublished = {Public Safety and Homeland Security Bureau, daily reports
                  26 September -- 19 October 2024},
  year         = {2024},
  note         = {Documents DOC-405827A1 through DOC-406771A1. Retrieved from
                  \url{https://docs.fcc.gov/public/attachments/}. The 24 daily
                  filings analysed here span 26 September -- 19 October 2024; the
                  26 September filing predates state-level cause reporting},
  url          = {https://www.fcc.gov/helene}
}

@misc{fcc_dirs_program,
  author       = {{Federal Communications Commission}},
  title        = {Disaster Information Reporting System ({DIRS})},
  year         = {2026},
  note         = {Voluntary reporting system activated for major disasters;
                  filings are aggregated and published daily},
  url          = {https://www.fcc.gov/general/disaster-information-reporting-system-dirs-0}
}

@misc{fcc_nors,
  author       = {{Federal Communications Commission}},
  title        = {Network Outage Reporting System ({NORS})},
  year         = {2026},
  note         = {Mandatory outage reporting under 47 C.F.R. Part 4;
                  filings are presumed confidential under 47 C.F.R. \S 4.2},
  url          = {https://www.fcc.gov/network-outage-reporting-system-nors}
}

@misc{ncsco2024october,
  author       = {Davis, Corey},
  title        = {October Dries Out in a Monthly Rainfall Reversal},
  howpublished = {North Carolina State Climate Office},
  month        = nov,
  year         = {2024},
  note         = {Published 4 November 2024. Statewide October 2024
                  precipitation averaged 0.54~in, the third-driest October
                  since 1895},
  url          = {https://climate.ncsu.edu/blog/2024/11/october-dries-out-in-a-monthly-rainfall-reversal}
}

@misc{domprep2024helene,
  author       = {Hauser, Greg},
  title        = {Bridging Communication Gaps: Lessons from {Hurricane Helene}},
  howpublished = {Domestic Preparedness Journal},
  month        = apr,
  year         = {2025},
  note         = {Published 30 April 2025. The author is the North Carolina
                  Division of Emergency Management statewide interoperability
                  coordinator and ESF-2 lead for the Helene response},
  url          = {https://domesticpreparedness.com/articles/bridging-communication-gaps-lessons-from-hurricane-helene}
}

@inproceedings{reed2018numerical,
  author    = {Reed, Dorothy A. and Wang, Shuoqi},
  title     = {Numerical Modeling of Power Delivery and Telecommunications
               Infrastructure for Hurricanes {H}arvey and {I}rma},
  booktitle = {Forensic Engineering 2018: Forging Forensic Frontiers},
  pages     = {1008--1016},
  publisher = {American Society of Civil Engineers},
  address   = {Austin, TX, USA},
  year      = {2018},
  doi       = {10.1061/9780784482018.097}
}

@article{claffy2022challenges,
  author  = {claffy, kc and Clark, David},
  title   = {Challenges in Measuring the {I}nternet for the Public Interest},
  journal = {Journal of Information Policy},
  volume  = {12},
  pages   = {195--233},
  year    = {2022},
  doi     = {10.5325/jinfopoli.12.2022.0003}
}

@article{feeny2026quantitative,
  author  = {Feeny, Nolan and White, Anna and Guikema, Seth},
  title   = {A Quantitative Spatial Approach to Estimate Cellular Network
             Coverage during Natural Hazards Using Publicly Available Data},
  journal = {Journal of Infrastructure Preservation and Resilience},
  volume  = {7},
  number  = {1},
  pages   = {14},
  year    = {2026},
  doi     = {10.1186/s43065-026-00176-0}
}

@misc{ncei_ghcnd,
  author       = {{NOAA National Centers for Environmental Information}},
  title        = {Global Historical Climatology Network Daily ({GHCN}-Daily),
                  accessed via the {NCEI} Access Data Service},
  year         = {2024},
  howpublished = {\url{https://www.ncei.noaa.gov/access/services/data/v1}},
  note         = {Dataset \texttt{daily-summaries}, element \texttt{PRCP};
                  retrieved for 31 stations in nine western North Carolina
                  counties, 13--15 October 2024}
}

@misc{ioda_api,
  author       = {{Georgia Institute of Technology Internet Intelligence Lab}},
  title        = {{IODA} v2 {API}: Internet Outage Detection and Analysis},
  year         = {2024},
  howpublished = {\url{https://api.ioda.inetintel.cc.gatech.edu/v2/}},
  note         = {Signal \texttt{ping-slash24}, region entities 4444
                  (North Carolina) and 4446 (Tennessee)}
}

@misc{fcc_dirs_irma,
  author       = {{Federal Communications Commission}},
  title        = {Communications Status Report for Areas Impacted by
                  {Hurricane Irma}},
  howpublished = {Public Safety and Homeland Security Bureau, report as of
                  11 September 2017, document DOC-346655A1},
  year         = {2017},
  url          = {https://docs.fcc.gov/public/attachments/DOC-346655A1.pdf}
}

@misc{fcc_dirs_maria,
  author       = {{Federal Communications Commission}},
  title        = {Communications Status Report for Areas Impacted by
                  {Hurricane Maria}},
  howpublished = {Public Safety and Homeland Security Bureau, report as of
                  23 September 2017, document DOC-346860A1},
  year         = {2017},
  url          = {https://docs.fcc.gov/public/attachments/DOC-346860A1.pdf}
}

@misc{fcc_dirs_ida,
  author       = {{Federal Communications Commission}},
  title        = {Communications Status Report for Areas Impacted by
                  {Hurricane Ida}},
  howpublished = {Public Safety and Homeland Security Bureau, report as of
                  30 August 2021, document DOC-375318A1},
  year         = {2021},
  url          = {https://docs.fcc.gov/public/attachments/DOC-375318A1.pdf}
}

@misc{fcc2013resiliency,
  author       = {{Federal Communications Commission}},
  title        = {Improving the Resiliency of Mobile Wireless Communications
                  Networks},
  howpublished = {Report and Order, PS Docket Nos.\ 13-239 and 11-60,
                  FCC 13-125},
  year         = {2013},
  note         = {Released 27 September 2013},
  url          = {https://docs.fcc.gov/public/attachments/FCC-13-125A1.pdf}
}

@article{menne2012ghcnd,
  author  = {Menne, Matthew J. and Durre, Imke and Vose, Russell S. and
             Gleason, Byron E. and Houston, Tamara G.},
  title   = {An Overview of the Global Historical Climatology Network-Daily
             Database},
  journal = {Journal of Atmospheric and Oceanic Technology},
  volume  = {29},
  number  = {7},
  pages   = {897--910},
  year    = {2012},
  doi     = {10.1175/JTECH-D-11-00103.1}
}

@article{gamboa2026helene,
  author  = {Gamboa, Stephen and Chelminski, Paul Roman and Shenvi, Christina},
  title   = {Lessons Learned From {Helene}: The Role of a Rural Community
             Hospital in Disaster Response After a Major Hurricane},
  journal = {Annals of Emergency Medicine},
  year    = {2026},
  doi     = {10.1016/j.annemergmed.2026.01.005},
  note    = {Disaster Medicine/Concepts section; published online
             10 February 2026, volume and pages not yet assigned}
}

@inproceedings{connelly2025radio,
  author    = {Connelly, Donald and Farmer, Betty and Spasovska, Katerina},
  title     = {The Role of Radio and Ham Radio When Everything Else Fails
               During Crisis: Lessons Learned From {Hurricane Helene}},
  booktitle = {International Crisis and Risk Communication Association Reports
               (2025 Annual Proceedings)},
  volume    = {13},
  number    = {1},
  pages     = {130--133},
  year      = {2025},
  doi       = {10.69931/001c.142861}
}
